\documentclass[twocolumn,showpacs,preprintnumbers,amsmath,amssymb]{revtex4}

\usepackage{graphicx}
\usepackage{dcolumn}
\usepackage{bm}

\usepackage{epsfig}

\begin{document}

\title{Stabilization of in-phase fluxon state by geometrical confinement in small  $\text{Bi}_{2}\text{Sr}_{2}\text{Ca}\text{Cu}_{2}\text{O}_{8+x}$ mesa structures.}

\author{S. O. Katterwe}
\author{V. M. Krasnov}
\email{Vladimir.Krasnov@physto.se}
\affiliation{
Department of Physics, Stockholm University, AlbaNova University Center, SE-10691 Stockholm, Sweden
}

\date{\today}

\begin{abstract}
The in-phase (rectangular) fluxon lattice is required for achieving coherent THz emission from stacked Josephson junctions. Unfortunately, it is usually unstable due to mutual repulsion of fluxons in neighbor junctions, which favors the out-of-phase (triangular) lattice. Here we experimentally study magnetic field modulation of the critical current in small Bi-2212 mesa structures with different sizes. A clear Fraunhofer-like modulation is observed when the field is aligned strictly parallel to superconducting CuO planes. For long mesas the periodicity of modulation is equal to half the flux quantum per intrinsic Josephson junction, corresponding to the triangular fluxon lattice. However, the periodicity is changed to one flux quantum, characteristic to the rectangular fluxon lattice, both by decreasing the length of the mesas and by increasing magnetic field. Thus, we demonstrate that the stationary in-phase fluxon state can be effectively stabilized by geometrical confinement in small Bi-2212 mesa structures.

\end{abstract}

\maketitle

Josephson flux-flow oscillators
can provide a remarkable linewidth $\sim 1$ Hz in the sub-THz frequency range \cite{Koshelets}, but with a small power $\lesssim \mu W$. 
The emission power can be increased by phase-locking of several coupled oscillators \cite{Barbara,SakUst}.
The strongest coupling is achieved between atomic scale intrinsic Josephson junctions (IJJs), naturally formed in single crystals of high temperature superconductor $\text{Bi}_{2}\text{Sr}_{2}\text{Ca}\text{Cu}_{2}\text{O}_{8+x}$ (Bi-2212) \cite{Kleiner94}. Furthermore, a large energy gap in Bi-2212 would allow operation in an important THz frequency range. Therefore, stacked IJJs are intensively studied as possible candidates for tunable, high-power THz oscillators \cite{SakUst,GrJensen,Machida,Koshelev,Batov,Ozyuzer,Bae2007}.

To achieve power amplification, junctions must be locked in the in-phase mode, so that Josephson vortices (fluxons) form a rectangular lattice in the stack. Unfortunately, the rectangular lattice is usually unstable due to mutual repulsion of fluxons, which favors formation of the triangular (out-of-phase) fluxon lattice. Motion of the triangular lattice leads to out-of-phase oscillations in neighbor junctions, which results in destructive interference and negligible emission. Thus, the major challenge for achieving coherent THz emission from stacked IJJ is to promote the in-phase fluxon state.

Fluxon distribution in stacked junctions is governed by three forces:
(i) the in-plane fluxon-fluxon repulsion; (ii) the in-plane fluxon-edge interaction 
and (iii) the interlayer fluxon-fluxon repulsion.
The latter is specific for stacks and promotes the triangular lattice. The two in-plane forces 
promote the rectangular fluxon lattice.
Therefore, the rectangular lattice can be stabilized via decreasing the interlayer force by decreasing the interlayer coupling \cite{KrasnovPhC98}, or via increasing the in-plane forces by reduction the junction length $L$ \cite{Koshelev,Machida,KrasnovPhC98} or by increasing the fluxon density \cite{Koshelev,Machida}. The latter can be described by the approximate inequality:
\begin{equation}\label{Eq1}
\Phi/\Phi_0 \gtrsim L/\lambda_J,
\end{equation}
where $\lambda_J$ the Josephson penetration depth, $\Phi=HLs$ is the flux through each IJJ, $s\simeq 1.5nm$ is the interlayer spacing of Bi-2212 and $\Phi_0=hc/2e$ is the flux quantum.

The symmetry of fluxon lattice can be revealed from modulation of the critical current as a function of in-plane magnetic field $I_c(H)$ \cite{Koshelev}: the rectangular lattice leads to conventional Fraunhofer $I_c(\Phi)$ modulation with $\Phi_0$ periodicity, while the triangular lattice leads to $\Phi_0/2$ periodicity because the fluxon lattice parameter in the $c-$axis direction is doubled. Previous measurements of $I_c(H)$ for IJJs didn't reveal clear periodic modulation \cite{Kleiner94,KrasnovPhC98,Latyshev,KrasnovComp,LatYamash}. The same is true even for low-$T_c$ stacks \cite{Kleiner94,KrasnovComp,stack}. This was explained by existence of multiple metastable fluxon modes in long, strongly coupled stacked junctions \cite{KrasnovPhC98,KrasnovComp}. So far, clear $\Phi_0/2$ and $\Phi_0$ modulations were observed only in the dynamic flux-flow state both for low-$T_c$ stacks \cite{KrasnovPRB96} and IJJs \cite{Ooi,WangFiske,KadowakiFF}. Yet, their interpretation is ambiguous because even the rectangular lattice may lead to the $\Phi_0/2$ modulation in the flux-flow state \cite{UstPed}. Therefore, it is necessary to analyze the {\it static} critical current, not the least because this is the most important fingerprint of the dc-Josephson effect.

Here we study size and magnetic field dependence of the critical current in small Bi-2212 mesa structures. We observe a clear Fraunhofer-like modulation of $I_{c}$ as a function of the in-plane magnetic field. The periodicity of modulation changes from half $\Phi_0$ - characteristic for the triangular lattice, to $\Phi_0$ - characteristic for the rectangular lattice, both upon increasing the field and decreasing the junction size. 
This provides a clear experimental evidence that the in-phase fluxon state can be stabilized by geometrical confinement in small stacked Josephson junctions, even in the stationary case.

Mesa structures, containing few IJJs, were fabricated on top of Bi-2212 single crystals with $T_{c} = 82$K. Some mesas were FIB-trimmed to reduce the size. Details of mesa fabrication can be found in Ref. \cite{KrSubmicron}. 
Table-\ref{tab:Properties} summarizes properties of studied mesas. Samples were mounted on a rotatable sample holder with alignment accuracy $\sim 0.02^{\circ}$. All measurements reported below were performed at $T=1.6$ K.

\begin{table}[t]
	\begin{ruledtabular}
		\begin{tabular}{ccccccc}
Mesa &$L$ [$\mu$m]&$W$ [$\mu$m]&$I_{c0}$ [$\mu$A]&$\lambda_J$ [$\mu$m]&$H_{0}$ [T]&$H_0^{exp}$ [T]\\
		\hline
4a& 5.1 & 1.4 & 75.0 & 0.68 & 0.27 & 0.26\\		
4b& 2.7 & 1.4 & 39.5 & 0.69 & 0.51 & 0.55\\
6b& 2.0 & 1.7 & 39.8 & 0.65 & 0.69 & 0.662\\
		\end{tabular}
		\end{ruledtabular}
	\caption{\label{tab:Properties}Properties of the measured mesas on the same Bi-2212 single crystal. $W$ is the width of the junctions parallel to the magnetic field, $\lambda_{J}=\sqrt{(\Phi_{0} c s LW)/(16 \pi^2 I_{c0} \lambda_{ab}^{2})}$, with $s = 1.5$ nm and $\lambda_{ab} = 200$ nm and $H_{0}=\Phi_0/Ls$, $H_0^{exp}$ is the measured periodicity of $I_c$-oscillations.}
\end{table}

In Bi-2212 mesas the top surface junction is always deteriorated and has
much smaller $I_c$ than the rest ``bulk" IJJs \cite{KrCondMat}. With increasing current, $I$, it  switches into the quasiparticle branch $V_1(I)$, while ``bulk" IJJs remain in the stationary superconducting state up to much larger ``bulk" critical current. Since we are interested in the collective behavior of the stack, associated with the symmetry of the fluxon lattice in the stack, we have to study the ``bulk" critical current. To do so, we first carefully measured the quasiparticle branch of the surface junction $V_1(I)$ at $H=0$. We then automatically subtracted it from the $I-V$ characteristics of the mesas.

\begin{figure}[t]
\includegraphics[width=3.0in]{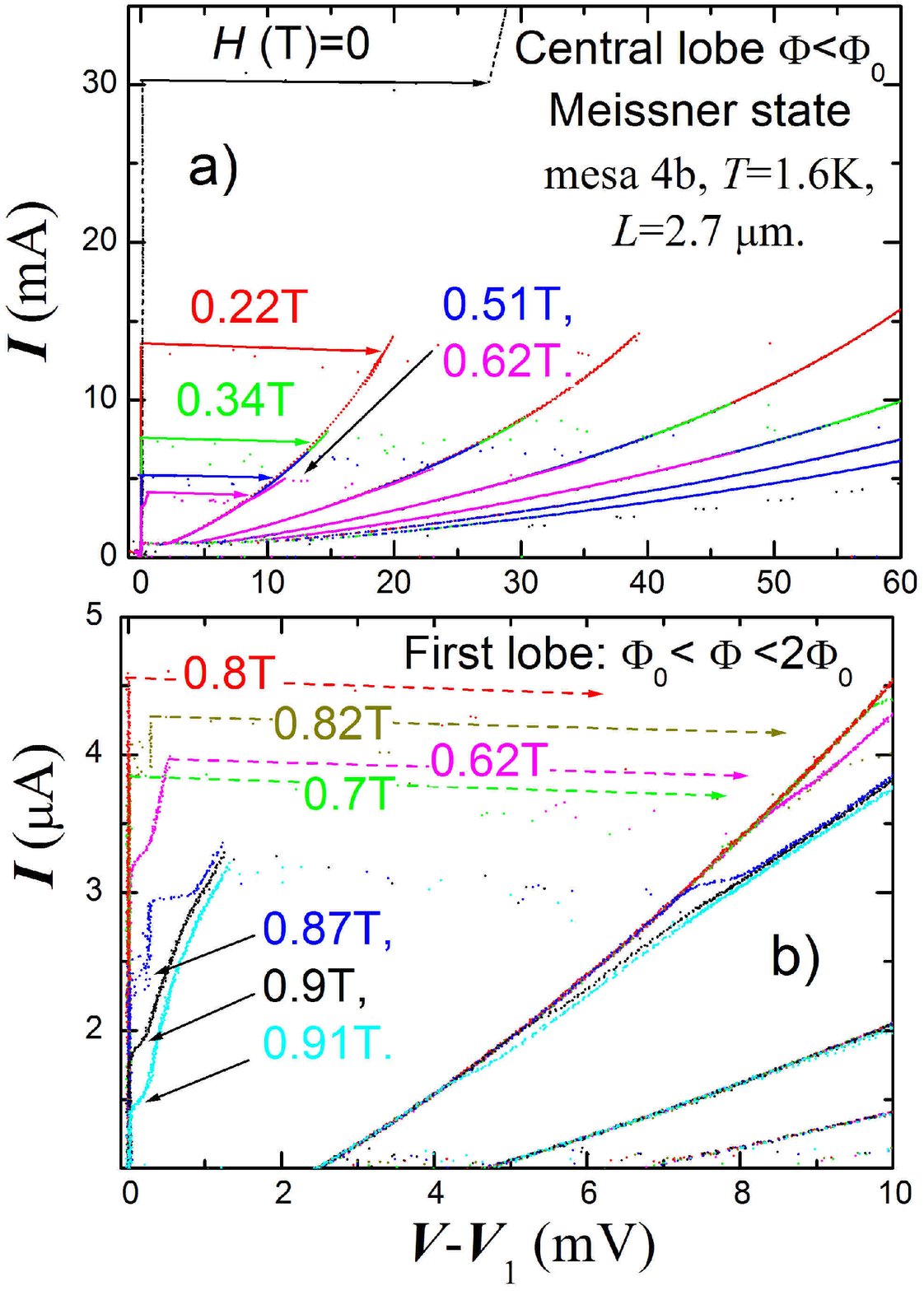}
\caption{\label{fig:iv} $I-V$ characteristics of mesa 4b with the subtracted quasiparticle branch of the surface junction, at different magnetic fields: a) within the central Meissner lobe of $I_c(\Phi)$ and b) for $1 <\Phi/\Phi_0 < 2$. Flux-flow and Fiske steps are seen.}
\end{figure}

Fig. \ref{fig:iv} shows current voltage characteristics of mesa 4b with the subtracted first quasiparticle branch $V_1$, for different magnetic fields. It is seen that subtraction works very well, with accuracy better than $10 \mu V$ at all fields. The bulk critical current $I_c$ was determined from the first deviation of $V-V_1$ from zero, using a $100 \mu$V criterion. A strong modulation of the critical current as a function of magnetic field is evident from Fig. \ref{fig:iv}. With increasing field, the flux-flow branch with distinct Fiske steps \cite{WangFiske} develops in the $I-V$ curve, as seen from Fig. \ref{fig:iv} b).

\begin{figure}[t]
\includegraphics[width=3.0in]{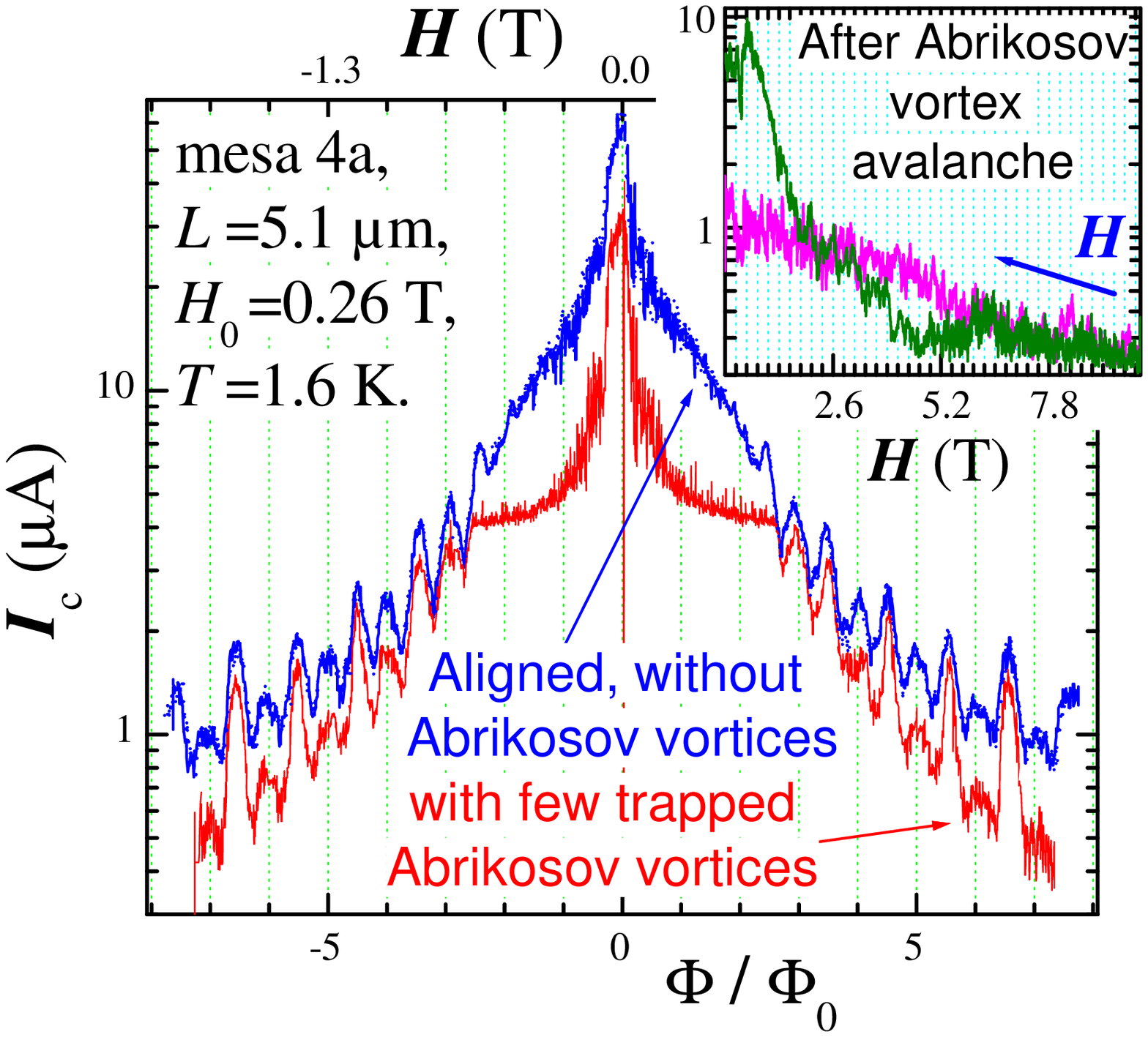} 
\caption{\label{fig:Ic} Experimental $I_c(\Phi)$ for the long mesa: without Abrikosov vortices (blue line, four field sweeps) and with few immobile Abrikosov vortices (red line, two field sweeps). Three field regions can be distinguished: i) $\Phi<2\Phi_0$: metastable region with strong fluctuations and without modulation; ii) $2\Phi_0<\Phi<5\Phi_0$: $\Phi_0/2$ modulation with even integer and half-integer maxima; iii) $\Phi>5\Phi_0$: transition to $\Phi_0$ modulation with predominance of half-integer maxima. Inset shows $I_c(\Phi)$ for two sweeps down from 10T.}
\end{figure}

Field alignment was crucial for observation of $I_c (H)$ modulation. Even slight misalignment with respect to CuO planes led to avalanche entrance of Abrikosov vortices with increasing field. After that the $I_c(H)$ patterns become heavily distorted, irreversible and completely lack periodic modulation, as shown in inset of Fig. \ref{fig:Ic}.
Thus, exact field alignment was the main experimental challenge in this work.

At low fields $H<6T$ the alignment was complicated by field lock-in \cite{KadowakiFF} along the CuO-plane, which led to a hysteresis of as much as three degrees, making the precise alignment impossible. To avoid the hysteresis, we looked at the high field $H=15$T magnetoresistance at bias close to the sum-gap voltage \cite{KrCondMat}. The latter is very sensitive to the c-axis field component, which leads to a substantial negative magnetoresistance \cite{MR}. To the contrary, the magnetoresistance was negligible for precise in-plane field alignment. Thus we were able to accurately align samples by minimizing the high field and high bias magnetoresistance. Even after perfect alignment, presence of trapped Abrikosov vortices affected the $I_c$, as shown by red line in Fig. 2. Yet, immobile Abrikosov vortices did not distort the qualitative shape of $I_c(H)$ modulation. To get rid of trapped Abrikosov vortices, samples were always heated up above $T_c$ after alignment.

\begin{figure}[t]
\includegraphics[width=3.0in]{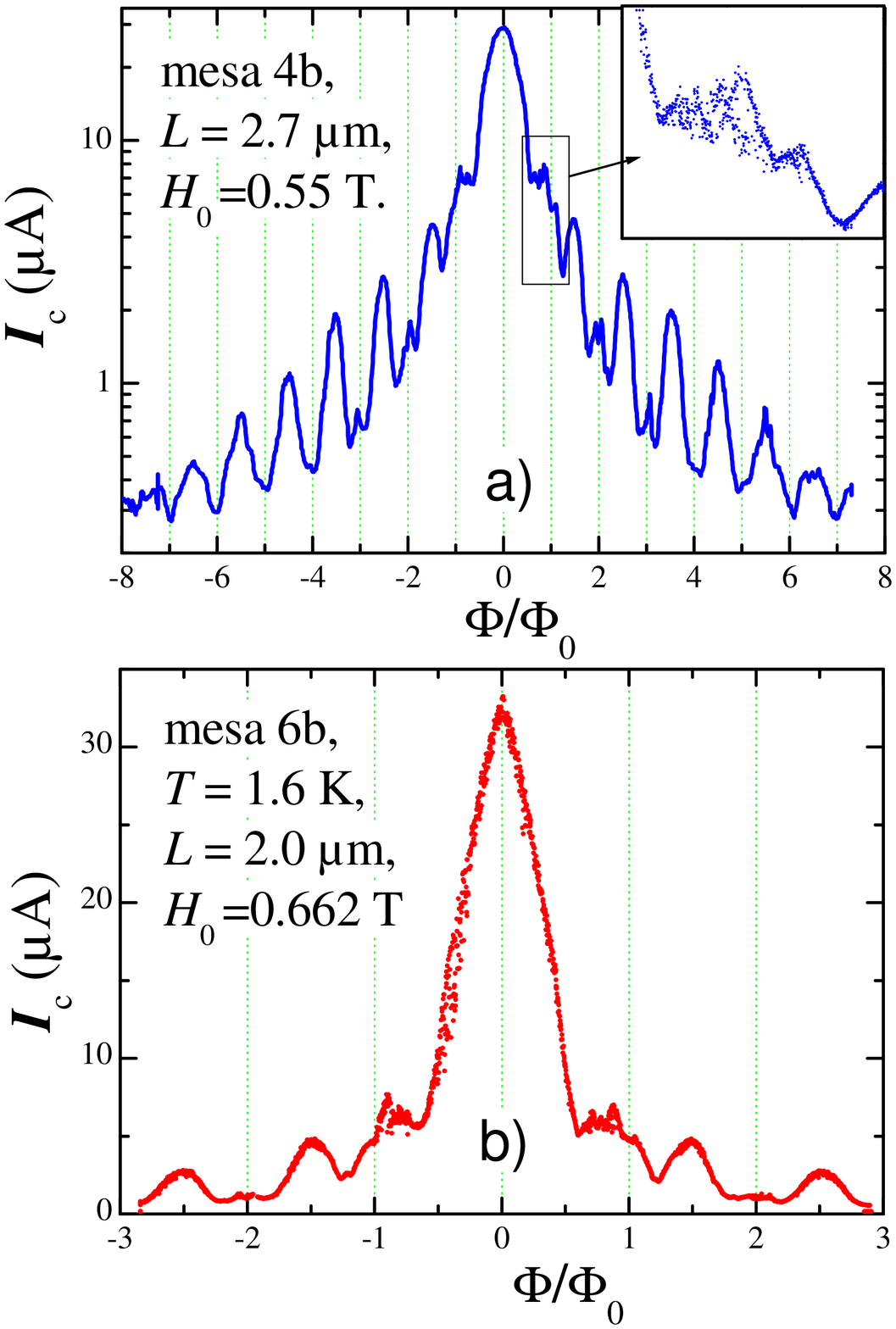} 
\caption{\label{fig:IcShort} Experimental $I_c(\Phi)$ for a) moderately long (in a semi-log scale) and b) the shortest mesas. It is seen that integer maxima are rapidly decreasing with increasing field and decreasing $L$. At $\Phi/\Phi_0 > L/\lambda_j$ only half-integer maxima are seen, indicating transition to the rectangular fluxon lattice. Inset in a) shows metastable sub-branches at $\Phi/\Phi_0 \sim 1$.}
\end{figure}

Figs. \ref{fig:Ic} and \ref{fig:IcShort} present the main experimental result of our work: magnetic field modulation of the ``bulk" critical current for Bi-2212 mesas of different sizes. Vertical grid lines correspond to integer flux quanta per IJJ.
All shown patterns are perfectly reversible. Magnetic field was swept back-and-forth several times (from -2 to +2 T in Figs. \ref{fig:Ic} and \ref{fig:IcShort} b), and from -4T to 4T in Fig. \ref{fig:IcShort} a) without significant changes in $I_c (H)$. This confirms that alignment was fine and Abrikosov vortices didn't enter the mesa up to the largest field.

Figure \ref{fig:Ic} a) shows the $I_c (\Phi)$ pattern (in a semi-log scale) for the long mesa (4a), with $L=5.1$ $\mu\text{m}$ and $L/\lambda_J \simeq 7.5$. When increasing the field from zero, $I_c$ first continuously decreasing and at $H \gtrsim 0.5$ T, the oscillatory behavior appears. The period of modulation is initially $\Phi_0/2$ rather than $\Phi_0$, with maxima both at integer and half-integer $\Phi_0$. However, with increasing field the amplitude of integer maxima rapidly decreases and at high fields maxima at half-integer $\Phi / \Phi_0$ become dominant.

Figure \ref{fig:IcShort} a) shows $I_c(\Phi)$ (in a semi-log scale) for the moderately long mesa 4b with $L=2.7 \mu$m and $L/\lambda_J \simeq 3.9$. Here maxima at integer $\Phi / \Phi_0$ are strongly suppressed even at low fields. At the condition of Eq.(\ref{Eq1}), $\Phi > 3 \Phi_0$, the integer maxima completely disappear and the $I_c(\Phi)$ switches to conventional Fraunhofer modulation with one flux-quantum periodicity and maxima at half-integer $\Phi / \Phi_0$.

Finally, $I_c(\Phi)$ for the shortest mesa (6b) is shown in figure \ref{fig:IcShort} b). The length of the mesa is $L=2.0$ $\mu\text{m}$ and $L/\lambda_J=3.1$. Here integer maxima are suppressed almost completely already for $\Phi = 2 \Phi_0$.

\begin{figure}[t]
\includegraphics[width=3.0in]{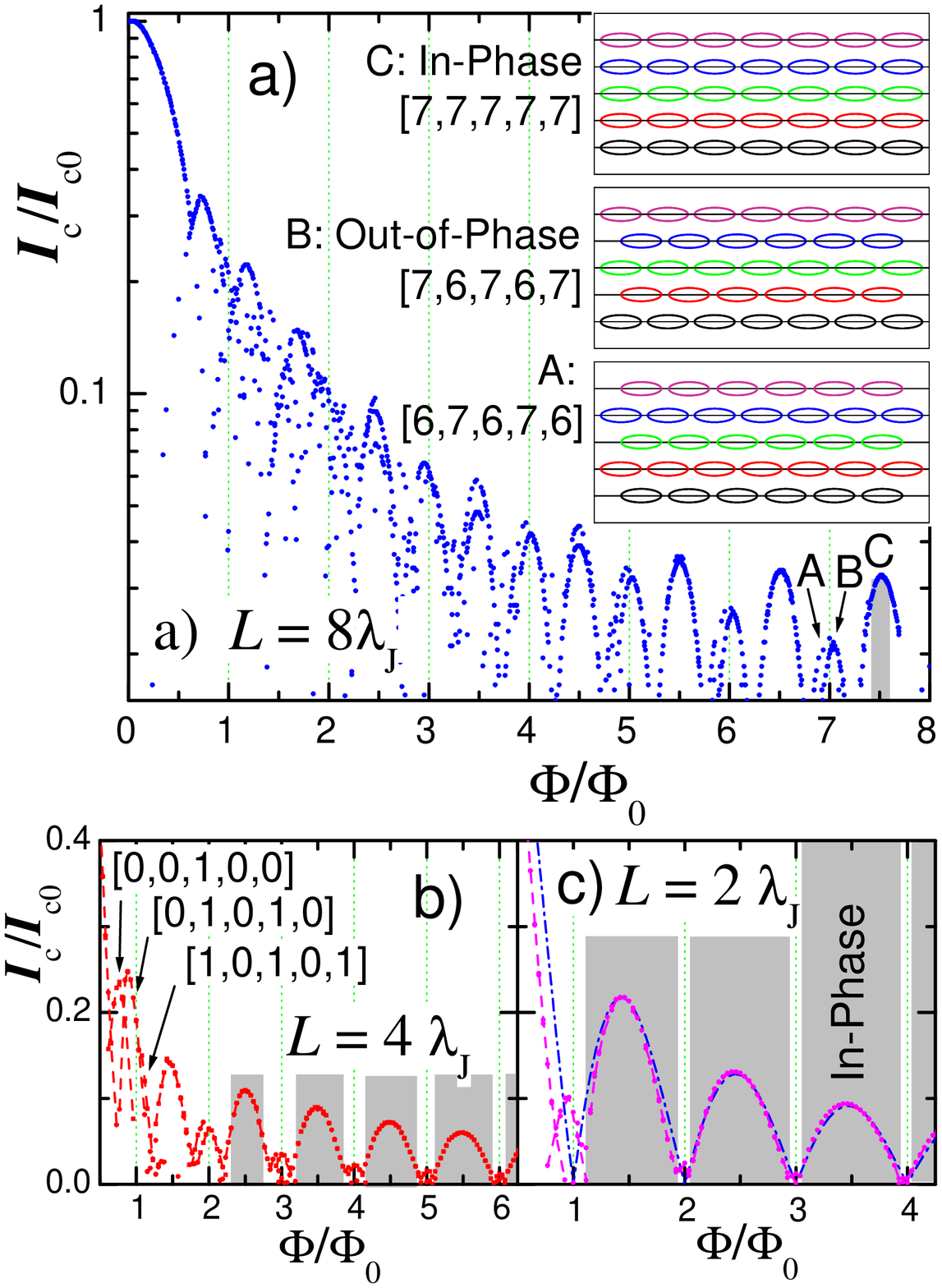}
\caption{\label{fig:Simul} Simulated $I_c(\Phi)$ for stack of $N=5$ IJJs for a) $L=8 \lambda_J$ (in a semi-log scale), b) $L=4 \lambda_J$ and c) $L=2 \lambda_J$. The transition from $\Phi_0/2$ to $\Phi_0$ periodicity occurs with increasing $\Phi$ and decreasing $L$ and reflects the transition from triangular to rectangular fluxon lattice, as shown by snapshots in panel a). Shaded areas indicate regions of stability of the rectangular (in-phase) lattice.
The in-phase mode is dominating in the shortest stack so that $I_c(\Phi)$ returns to the conventional Fraunhofer pattern, shown by the blue dashed-dotted line.
}
\end{figure}


For better understanding of experimental data, in Fig. \ref{fig:Simul} we show numerically simulated $I_c(\Phi)$ for $N=5$ stacked IJJs of different lengths: a) $L=8\lambda_J$ (in the semi-log scale), b) $L=4\lambda_J$ and c) $L=2\lambda_J$ (for details see the Supplementary \cite{Supplem}). It is seen that the overall behavior of $I_c(\Phi,L)$ is very similar to experimental data in Figs. \ref{fig:Ic} and \ref{fig:IcShort}:

For the long stack in Fig. 4 a) three field regions can be distinguished, similar to that in Fig. 2:

i) At low field $\Phi \lesssim 2 \Phi_0$ there are strong fluctuations of $I_c$ due to chaotic switching between multiple metastable fluxon modes \cite{KrasnovPhC98,KrasnovComp}, and no clear modulation of $I_c(\Phi)$.

ii) At intermediate fields $2 \Phi_0 <\Phi < 6 \Phi_0$ modulation with periodicity of $\Phi_0/2$ appears. The fluxon lattice is not yet formed in this field range and modulation is due to switching between certain quasi-periodic modes \cite{Supplem}.

iii) Finally at high field, approximately determined by condition of Eq. (\ref{Eq1}), the regular fluxon lattice is formed and maxima at integer $\Phi/\Phi_0$, corresponding to the triangular lattice become weaker than maxima at half-integer $\Phi/\Phi_0$, corresponding to the rectangular lattice, so that periodicity of modulation becomes $\Phi_0$.

In shorter stacks, the metastable region becomes smaller. In Figs. 3 a) and 4 b) it is seen only at $\Phi \sim \Phi_0$. In inset of Fig. 3 a) we show this metastable region, measured by repeated field sweeps. Certain subbranches can be distinguished, indicating a reduced variety of metastable fluxon modes in comparison to the long junctions case. In Figs. 4 b) we marked the most probable modes, which cause distinct subbranches in $I_c(\Phi)$. The metastable states gradually die out with decreasing $L$, as seen from the data for the shortest junctions in Figs. 3 b) and 4 c), and eventually disappear for $L < \lambda_J$.

Even region-ii) disappears in shorter stacks and the amplitude of integer maxima is always much  smaller than of half-integer maxima. The integer maxima are further suppressed with increasing field, see Figs. 4 b) and 3 a) and with decreasing junction length, see Figs. 4 c) and 3 b). For shorter junctions the $\Phi_0$ periodicity is established at smaller $\Phi/\Phi_0$, in agreement with Eq.(\ref{Eq1}). In short stacks, $L < \lambda_J$, only maxima at half-integer $\Phi/\Phi_0$ are left, and $I_c(\Phi)$ returns to the conventional Fraunhofer pattern, shown by the blue dashed-dotted line in Fig. 4 c). The apparent similarity between measured and simulated data in Figs. 2-3 confirms our estimation of $\lambda_J\sim 0.7 \mu m$ in Table- \ref{tab:Properties}.

The observed transition from $\Phi_0/2$ to $\Phi_0$ periodicity of $I_c(\Phi)$ modulation is the consequence of transition from triangular to rectangular fluxon lattice in the stack \cite{Machida,Koshelev}. This is illustrated by snapshots in Fig. 4 a), which show fluxon distributions at both sides of the subdominant maximum at $\Phi/\Phi_0 \simeq 7$ and the dominant maximum at $\Phi/\Phi_0 \simeq 7.5$. It is seen that the rectangular (in-phase) lattice occurs at the dominant half-integer maximum. For the stack with $L=2\lambda_J$ the in-phase state is stable practically in the whole field range, marked by shaded area in Fig. 4. All this could be verified using the real-time simulation code provided in the Supplementary \cite{Supplem}.

In summary, we observed clear Fraunhofer-like modulation of $I_c(H)$ in Bi-2212 mesa of different sizes. We demonstrated that the periodicity of modulation changes from half-flux quantum to flux quantum both with increasing magnetic field and decreasing mesa size. This indicates transition from triangular (out-of-phase) to rectangular (in-phase) fluxon lattice in stacked intrinsic Josephson junctions. Therefore, we confirm experimentally that the stationary in-phase fluxon state can be stabilized by geometrical confinement in small Bi-2212 mesa structures. This is important for realization of high power flux-flow oscillator in the THz frequency range.

We are grateful to A.Tkalecz and A.Rydh for assistance in experiment. The work was supported by K. \& A. Wallenberg foundation, the Swedish Research Council and the SU-Core Facility in Nanotechnology.



\end {document}